\documentstyle[12pt,qqa4lart]{article}
\input tcilatex

\QQQ{Language}{
American English
}

\begin{document}

\author{Lu-Ming Duan\thanks{%
correspondence author} and Guang-Can Guo\thanks{%
Electronic address: gcguo@sunlx06.nsc.ustc.edu.cn} \\
Department of Physics and Nonlinear Science Center,\\
University of Science and Technology of China,\\
Hefei 230026, People's Republic of China}
\title{Pulse Controlled Noise Suppressed Quantum Computation}
\date{}
\maketitle

\begin{abstract}
\baselineskip 24ptTo make arbitrarily accurate quantum computation possible,
practical realization of quantum computers will require suppressing noise in
quantum memory and gate operations to make it below a threshold value. A\
scheme based on realistic quantum computer models is described for
suppressing noise in quantum computation without the cost of stringent
quantum computing resources.
\end{abstract}

\newpage\baselineskip 24ptQuanum computation has become a very active field
ever since the discovery that quantum computers can be much more powerful
than their classical counterparts [1-3]. Quantum computers act as
sophisticated quantum information processors, in which calculations are made
by the controlled time evolution of a set of coupled two-level quantum
systems (qubits). Coherence in the evolution is essential for taking
advantage of quantum parallelism. However, there is a major obstacle to
realization of quantum computation. Decoherence of the qubits caused by the
inevitable interaction with noisy environment will make quantum information
too fragile to be of any practical use. Recently, interest in quantum
computation has increased dramatically because of two respects of advances
toward overcoming the above difficulty. First, a combination of a series of
innovative discoveries, such as quantum error correcting codes [4,5],
fault-tolerant error correction techniques [6], and concatenated coding [7],
has yielded the important threshold result [7,8], which promises that
arbitrarily accurate quantum computation is possible provided that the error
per operation is below a threshold value. Hence, noise below a certain level
is not an obstacle to reliable quantum computation. Second, great progress
has been made toward building the necessary quantum hardwares [9-11]. One
particularly attractive physical system is the nuclear spin because of its
good isolation from noisy environment. Some simple but real quantum
computations have been demonstrated in bulk spin resonance quantum systems
[12,13]; and a radical scheme, using semiconductor physics to manipulate
nuclear spins, has been recently proposed, which indicates a promising route
to large-scale quantum computation [14].

Fault-tolerant quantum error correction schemes are effective only when the
error rate per operation is below a threshold value. Operations in quantum
computers include transmission or storage of quantum states, and quantum
logic. For quantum logic, the estimated threshold error rate is about $%
10^{-6}$ [7,8]. Even for the most promising quantum hardware, the nuclear
spin system, the real noise seems also necessarily beyond this threshold
with the present technology [8]. Therefore, it is essentially important to
further suppress noise before taking the procedure of fault-tolerant quantum
error correction. Here, we propose a noise suppression scheme which operates
by applying a sequence of bit-flipping or phase-flipping pulses on the
system, with the pulse period less than the noise correlation time. It is
shown that noise in the pulse-controlled system can be greatly reduced.

The scheme based on pulse control is relatively easy to implement in
experiments. In fact, clever pulse methods (called refocusing techniques)
have been developed for years in nuclear magnetic resonance spectroscopy
(NMR) to eliminate many kinds of classical dephasing effects [15]. In a
recent work [16], Viola and Lloyd first used this technique to combat
quantum noise in computer memory with a specific single-qubit dephasing
model. In this paper, we consider the most general type of environmental
noise, including classical or quantum dephasing and dissipation as its
special case. By applying suitable pulse sequences, it is found that the
noise in the controlled system is much reduced; and more importantly, the
scheme can be readily extended to suppress noise in quantum gates (mainly
referring to two-bit quantum gates, whose error rate is much larger than
that of single-qubit rotations [9,14,17]). Suppose $p_g$ is the error rate
per operation, and $p_0$ is the additional error rate introduced by each
pulse. Our result shows that the error rate per operation in the
pulse-controlled system approximately reduces to $p_gp_0t_{dec}/t_c$, where $%
t_{dec}$ and $t_c$ are respectively the decoherence time and the noise
correlation time. Since $p_0$ can be made very small, the reducing factor $%
p_0t_{dec}/t_c$ is much less than $1$ even that the noise correlation time
is considerably smaller than the decoherence time. Therefore, by this scheme
it is possible for the reduced error rate to attain the threshold value,
say, $10^{-6}$. This is a desirous result for reliable quantum computation.
The only requirement in our scheme is that the pulses should be applied
frequently so that the pulse period is less than the noise correlation time.
This is experimentally feasible [11,15,16]. The scheme costs no additional
quantum computing resources.

First we show how to use pulse control to combat decoherence of a single
qubit in quantum memory. The qubit is described by Pauli's operator $%
\overrightarrow{\sigma }$. In the interaction picture, the most general form
of the Hamiltonian describing single-qubit decoherence due to environmental
noise can be expressed as (setting $\hbar =1$) 
\begin{equation}
\label{1}H_I\left( t\right) =\stackunder{\alpha =x,y,z}{\sum }\left[ \sigma
^\alpha \Gamma _\alpha \left( t\right) \right] ,
\end{equation}
where $\Gamma _\alpha \left( t\right) $, generally independent of time, are
noise terms, which may be classical stochastic variables or stochastic
quantum operators, corresponding classical noise or quantum noise ,
respectively. For environmental noise in quantum computers, it is reasonable
to assume that $\Gamma _\alpha \left( t\right) $ $\left( \alpha
=x,y,z\right) $ satisfy the conditions $\left\langle \Gamma _\alpha \left(
t\right) \right\rangle _{env}=0$ and 
\begin{equation}
\label{2}\left\langle \Gamma _\alpha \left( t\right) \Gamma _\beta \left(
t^{^{\prime }}\right) \right\rangle _{env}=f_{\alpha \beta }\left( \xi
\right) ,
\end{equation}
where $\left\langle \cdots \right\rangle _{env}$ denotes average over the
environment. In Eq. (2), $f_{\alpha \beta }\left( \xi \right) $ $\left(
\alpha ,\beta =x,y,z\right) $ are correlation functions and $\xi =\frac{%
t-t^{^{\prime }}}{t_c}$. The quantity $t_c$ characterizes the order of
magnitude of noise correlation times. Different noise correlation terms may
have different correlation times. But we assume that they have the same
order of magnitude, which is denoted by $t_c$. For simplicity, in the
following we directly call $t_c$ the noise correlation time.

Suppose that the qubit is initially in a pure state $\left| \Psi \left(
0\right) \right\rangle $. Under the Hamiltonian (1), after a short time $t$
it evolves into the following mixed state 
\begin{equation}
\label{3}\rho \left( t\right) =\left\langle T\left\{ e^{-i\int_0^tH_I\left(
t^{^{\prime }}\right) dt^{^{\prime }}}\right\} \left| \Psi \left( 0\right)
\right\rangle \left\langle \Psi \left( 0\right) \right| T\left\{
e^{i\int_0^tH_I\left( t^{^{\prime }}\right) dt^{^{\prime }}}\right\}
\right\rangle _{env},
\end{equation}
where $T\left\{ \cdots \right\} $ indicates that time-ordered product is
taken in the bracket. The difference between the states $\rho \left(
t\right) $ and $\left| \Psi \left( 0\right) \right\rangle $ stands for
errors. The error rate $p$ can be described by $p=1-F\left( t\right) $,
where $F\left( t\right) $ is the input-output state fidelity, defined as $%
F\left( t\right) =\left\langle \Psi \left( 0\right) \right| \rho \left(
t\right) \left| \Psi \left( 0\right) \right\rangle $. From Eqs. (1) (2) and
(3), it is not difficult to obtain an explicit perturbative expression for
the error rate. Up to the second order of the Hamiltonian, the result is 
\begin{equation}
\label{4}p=2\stackunder{\alpha ,\beta }{\sum }\left\langle \Delta \sigma
^\alpha \Delta \sigma ^\beta \right\rangle _s\int_0^t\int_0^{t_1}f_{\alpha
\beta }\left( \frac{t_2}{t_c}\right) dt_2dt_1,
\end{equation}
where $\Delta \sigma ^\alpha =\sigma ^\alpha -\left\langle \sigma ^\alpha
\right\rangle _s$, and $\left\langle \cdots \right\rangle _s$ denotes
average over the system. Equation (4) is derived with a pure input state.
However, it remains true when the qubit is initially in a mixed state. In
this case, we define the error rate by $p=1-F_e\left( t\right) $, where $%
F_e\left( t\right) $ is the entanglement fidelity [18], a natural extension
of the input-output fidelity to the mixed state circumstance. With this
definition, the expression for the error rate remains completely same as Eq.
(4).

Now we show how to use pulse control to reduce the error rate in quantum
memory. We apply a sequence of bit-flipping or phase-flipping pulses on the
qubit, with the pulse period $t_\Delta $ and pulse width $t_w$. It is
required that $t_w<<t_\Delta $ so that the environment-induced system
evolution during the short time $t_w$ is negligible. At time $t=0$, no pulse
is applied, and its operation is represented by the unit operator $I$. At
time $t=t_\Delta $, we begin to apply in turn the bit-flipping and the
phase-flipping pulses. The pulse-induced operations are thus respectively $%
I,\sigma ^x,\sigma ^z,\sigma ^x,\sigma ^z,\cdots $. Four pulse periods make
up a control period. In the $(n+1)$th control period (form time $4nt_\Delta $
to time $4\left( n+1\right) t_\Delta $), the effective system evolution
under pulse control is represented by the following evolution operator 
\begin{equation}
\label{5}
\begin{array}{c}
U_{n+1}=\sigma ^zT\left\{ \exp \left[ -i\int_{\left( 4n+3\right) t_\Delta
}^{\left( 4n+4\right) t_\Delta }H_I\left( t^{^{\prime }}\right) dt^{^{\prime
}}\right] \right\} \sigma ^xT\left\{ \exp \left[ -i\int_{\left( 4n+2\right)
t_\Delta }^{\left( 4n+3\right) t_\Delta }H_I\left( t^{^{\prime }}\right)
dt^{^{\prime }}\right] \right\}  \\  
\\ 
\times \sigma ^zT\left\{ \exp \left[ -i\int_{\left( 4n+1\right) t_\Delta
}^{\left( 4n+2\right) t_\Delta }H_I\left( t^{^{\prime }}\right) dt^{^{\prime
}}\right] \right\} \sigma ^xT\left\{ \exp \left[ -i\int_{4nt_\Delta
}^{\left( 4n+1\right) t_\Delta }H_I\left( t^{^{\prime }}\right) dt^{^{\prime
}}\right] \right\}  \\  
\\ 
=-T\left\{ \exp \left[ -i\int_{\left( 4n+3\right) t_\Delta }^{\left(
4n+4\right) t_\Delta }\sigma ^y\sigma ^xH_I\left( t^{^{\prime }}\right)
\sigma ^x\sigma ^ydt^{^{\prime }}-i\int_{\left( 4n+2\right) t_\Delta
}^{\left( 4n+3\right) t_\Delta }\sigma ^yH_I\left( t^{^{\prime }}\right)
\sigma ^ydt^{^{\prime }}\right. \right.  \\  
\\ 
\left. \left. -i\int_{\left( 4n+1\right) t_\Delta }^{\left( 4n+2\right)
t_\Delta }\sigma ^xH_I\left( t^{^{\prime }}\right) \sigma ^xdt^{^{\prime
}}-i\int_{4nt_\Delta }^{\left( 4n+1\right) t_\Delta }H_I\left( t^{^{\prime
}}\right) dt^{^{\prime }}\right] \right\}  \\  
\\ 
=-T\left\{ \exp \left[ -i\int_{4nt_\Delta }^{4\left( n+1\right) t_\Delta }%
\stackunder{\alpha =x,y,z}{\sum }\left[ \sigma ^\alpha \Gamma _\alpha
^{^{\prime }}\left( t^{^{\prime }}\right) \right] dt^{^{\prime }}\right]
\right\} ,
\end{array}
\end{equation}
where $T\left\{ \cdots \right\} $ denotes the time-ordered product. The
effective noise terms $\Gamma _\alpha ^{^{\prime }}\left( 4nt_\Delta
+t\right) $ with $\alpha =x,y,z$ and $0\leq t<4t_\Delta $ in Eq. (5) are
defined respectively as follows 
\begin{equation}
\label{6}
\begin{array}{c}
\Gamma _x^{^{\prime }}\left( 4nt_\Delta +t\right) =\frac 14\left[ \Gamma
_x\left( 4nt_\Delta +\frac t4\right) +\Gamma _x\left( 4nt_\Delta +t_\Delta
+\frac t4\right) \right.  \\  
\\ 
\left. -\Gamma _x\left( 4nt_\Delta +2t_\Delta +\frac t4\right) -\Gamma
_x\left( 4nt_\Delta +3t_\Delta +\frac t4\right) \right] ,
\end{array}
\end{equation}
\begin{equation}
\label{7}
\begin{array}{c}
\Gamma _y^{^{\prime }}\left( 4nt_\Delta +t\right) =\frac 14\left[ \Gamma
_y\left( 4nt_\Delta +\frac t4\right) -\Gamma _y\left( 4nt_\Delta +t_\Delta
+\frac t4\right) \right.  \\  
\\ 
\left. +\Gamma _y\left( 4nt_\Delta +2t_\Delta +\frac t4\right) -\Gamma
_y\left( 4nt_\Delta +3t_\Delta +\frac t4\right) \right] ,
\end{array}
\end{equation}
\begin{equation}
\label{8}
\begin{array}{c}
\Gamma _z^{^{\prime }}\left( 4nt_\Delta +t\right) =\frac 14\left[ \Gamma
_z\left( 4nt_\Delta +\frac t4\right) -\Gamma _z\left( 4nt_\Delta +t_\Delta
+\frac t4\right) \right.  \\  
\\ 
\left. -\Gamma _z\left( 4nt_\Delta +2t_\Delta +\frac t4\right) +\Gamma
_z\left( 4nt_\Delta +3t_\Delta +\frac t4\right) \right] ,
\end{array}
\end{equation}
After pulse control, the only difference in the system evolution (5) is that
the noise terms $\Gamma _\alpha \left( t\right) $ are replaced by the
corresponding effective noise terms $\Gamma _\alpha ^{^{\prime }}\left(
t\right) $. Form Eqs. (2) and (6-8), it is not difficult to get the
correlations of the effective noise $\Gamma _\alpha ^{^{\prime }}\left(
t\right) $. Then, substituting these correlations into Eq. (4), we obtain
the error rate $p_c$ after pulse control. The final result is 
\begin{equation}
\label{9}\frac{p_c}p=\alpha \left( \frac{t_\Delta }{t_c}\right) ^2,
\end{equation}
where $t_c$ is the noise correlation time. The normalized constant $\alpha $
can be approximated by the following expression 
\begin{equation}
\label{10}\alpha \simeq \frac{\stackunder{\alpha ,\beta }{\sum }\left\langle
\Delta \sigma ^\alpha \Delta \sigma ^\beta \right\rangle
_s\int_0^t\int_0^{t_1/t_c}\Delta _\alpha \Delta _\beta \frac{\partial ^2}{%
\partial \xi ^2}f_{\alpha \beta }\left( \xi \right) d\xi dt_1}{\stackunder{%
\alpha ,\beta }{\sum }\left\langle \Delta \sigma ^\alpha \Delta \sigma
^\beta \right\rangle _s\int_0^t\int_0^{t_1/t_c}f_{\alpha \beta }\left( \xi
\right) d\xi dt_1},
\end{equation}
where $\Delta _x\simeq -1,$ $\Delta _y\simeq -\frac 12,$ and $\Delta
_z\simeq o\left( t_\Delta /t_c\right) .$ The symbol $o\left( t_\Delta
/t_c\right) $ indicates that $\Delta _z$ and $t_\Delta /t_c$ have the same
order of magnitude. By this notation, we have $\alpha \simeq o\left(
1\right) $. The factor $\alpha $ is unimportant to our result and will be
omitted in the following discussion. From Eq. (10), it follows that through
pulse control the error rate in quantum memory can be reduced by a factor
proportional to the second order of the rate of the pulse period to the
noise correlation time.

In fact, by applying more complicate pulse sequences, the error rate can be
further reduced. For example, we may apply the pulse sequence $I,\sigma
^x,\sigma ^z,\sigma ^x,I,\sigma ^x,\sigma ^z,$ $\sigma ^x,I,\cdots $, with
the pulse period $t_\Delta $. In this sequence, a control period consists of
eight pulse periods. We can similarly calculate the error rate of the system
controlled by this pulse sequence. The result is that the error rate is
reduced by a factor proportional to $\left( t_\Delta /t_c\right) ^4$. In
general, if we apply a pulse sequence with the control period consisting of $%
2^{n+1}$ pulse periods, the error rate is able to be reduced by a factor
proportional to $\left( t_\Delta /t_c\right) ^{2n}$.

In the above, we considered decoherence of a single qubit. Now, Suppose that
there are $L$ qubits, described respectively by Pauli's operators $%
\overrightarrow{\sigma }_l$. In the interaction picture, the Hamiltonian
describing decoherence of $L$ qubits can be generally expressed as 
\begin{equation}
\label{11}H_L\left( t\right) =\stackunder{l}{\sum }\stackunder{\alpha =x,y,z%
}{\sum }\left[ \sigma _l^\alpha \Gamma _l^\alpha \left( t\right) \right] ,
\end{equation}
where the noise terms satisfy the conditions $\left\langle \Gamma _l^\alpha
\left( t\right) \right\rangle _{env}=0$ and 
\begin{equation}
\label{12}\left\langle \Gamma _l^\alpha \left( t\right) \Gamma _{l^{^{\prime
}}}^\beta \left( t^{^{\prime }}\right) \right\rangle _{env}=f_{ll^{^{\prime
}}}^{\alpha \beta }\left( \frac{t-t^{^{\prime }}}{t_c}\right) .
\end{equation}
The above description of many qubit decoherence is quite general. It
includes independent decoherence and cooperative decoherence as its special
case [19,20]. For reducing decoherence of this system, the control method is
very simple. We need only apply the above pulse sequence separately on each
qubit. It is easy to know that the error rate of the controlled system will
be reduced by a factor proportional to $\left( t_\Delta /t_c\right) ^{2n}$,
where $n$ depends on which pulse sequence is chosen.

To make fault-tolerant quantum computation possible, noise in quantum gates
should be suppressed as well as noise in quantum memory. The pulse control
scheme can be readily extended to include quantum gate operations. Any
quantum gate can be decomposed into a series of quantum controlled-NOT
(CNOT) gates together with some single-qubit rotations [21]. In general, the
time required to perform quantum CNOT is much larger than the running time
of single qubit rotations [9,14,17]. Therefore, environmental noise has much
more influence on the quantum CNOT\ than on single-qubit rotations. To make
fault-tolerant quantum computation possible, the key step is thus to
suppress environmental noise in quantum CNOT\ gates. The quantum CNOT logic
makes use of direct or indirect interaction between the qubits. The
interaction is usually described by the following effective Hamiltonian
[11,14] 
\begin{equation}
\label{13}H_g=\stackunder{ll^{^{\prime }}}{\sum }g_{ll^{^{\prime }}}\left(
t\right) \overrightarrow{\sigma }_l\cdot \overrightarrow{\sigma }%
_{l^{^{\prime }}},
\end{equation}
where $g_{ll^{^{\prime }}}\left( t\right) $, possibly independent of time,
are coupling coefficients. The environmental noise in these gates can be
easily suppressed by the pulse control method. We need only synchronize the
pulses acting on different qubits. The synchronized pulses induce either a
collective bit flip $\otimes _l\sigma _l^x$ or a collective phase flip $%
\otimes _l\sigma _l^z$. The gate Hamiltonian remains unchanged under these
operations, i.e., 
\begin{equation}
\label{14}\left( \otimes _l\sigma _l^x\right) H_g\left( \otimes _l\sigma
_l^x\right) =\left( \otimes _l\sigma _l^z\right) H_g\left( \otimes _l\sigma
_l^z\right) =H_g.
\end{equation}
Hence the quantum CNOT logic is not influenced by the pulses, whereas the
environmental noise (described by the Hamiltonian (11)) is greatly
suppressed.

We have shown that environmental noise, whether in quantum memory or in
quantum gates, all can be reduced by the pulse control method. In the
analyses, we make an ideal assumption that the pulses introduce no
additional noise. If the inaccuracy of the pulse is considered, an
additional error rate $p_0$ will be introduced by each pulse. Now we
estimate in this circumstance to what amount the error rate per quantum
operation can be reduced through pulse control. Let $t_g$ and $p_g$ denote
the running time and the environment-induced error rate of the operation,
respectively, then $t_g/t_\Delta $ defines the number of pulses applied
during the operation. After pulse control, the error rate approximately
becomes 
\begin{equation}
\label{15}p_g^{^{\prime }}\simeq p_g\left( \frac{t_\Delta }{t_c}\right)
^{2n}+\frac{t_g}{t_\Delta }p_0\geq p_g\left( 2n+1\right) \left( \frac{p_0t_g%
}{2np_gt_c}\right) ^{\frac{2n}{2n+1}},
\end{equation}
where the minimum is attained when $t_\Delta =\left( \frac{p_0t_gt_c^{2n}}{%
2np_g}\right) ^{\frac 1{2n+1}}$, which is the optimal value for the pulse
period. The parameter $n$ in Eq. (15) depends on which pulse sequence is
chosen. If $2n$ is considerably large, $p_g^{^{\prime }}/p_g\simeq \left(
p_0t_g\right) /\left( p_gt_c\right) $. For environment-induced error, $p_g$
can be approximated by $p_g\simeq t_g/t_{dec}$ [17], where $t_{dec}$ is the
decoherence time. With this approximation, we have $p_g^{^{\prime
}}/p_g\simeq p_0t_{dec}/t_c$. The error rate is reduced by a very small
factor $p_0t_{dec}/t_c$. This suggests that the pulse control method is very
effective.

Compared with other noise suppression schemes, the pulse control method has
several remarkable features. First, it costs no additional qubits. This is
an important feature since with the present technology, quantum computing
resources are still very stringent [10,11,22]. Only small quantum systems
have been demonstrated experimentally. For these systems, it is impossible
to perform fault-tolerant quantum error correction, but the pulse control
scheme works well. Second, in the pulse control scheme, no encoding,
decoding, and error correction are required, and no measurement is
performed. Hence, in contrast to quantum error correction, this scheme
introduces no slowdown of the computation speed. Last but not the least,
pulse control is relatively a mature technology in experiments, especially
in the NMR systems [15]. The pulse control method, combined with
fault-tolerant quantum error correction, may ultimately make reliable
quantum computation possible.\\

{\bf Acknowledgment}

This project was supported by the National Natural Science Foundation of
China.

\newpage\


\begin{thebibliography}{99}
\bibitem{1}  P. W. Shor, in Proc. of the 35th Annual Symposium on
Foundations of Computer Science (IEEE Press, Los Alamitos, CA, 1994),
pp.124-134.

\bibitem{2}  S. Lloyd, Science 273, 1073 (1996).

\bibitem{3}  L. K. Grover, Phys. Rev. Lett. 79, 325 (1997).

\bibitem{4}  P. W. Shor, Phys. Rev. A 52, R2493 (1995).

\bibitem{5}  A. M. Steane, Phys. Rev. Lett. 77, 793 (1996).

\bibitem{6}  D. P. DiVincenzo and P. W. Shor, Phys. Rev. Lett. 77, 3260
(1996).

\bibitem{7}  E. Knill, R. Laflamme, and W. H. Zurek, Science 279, 342 (1998).

\bibitem{8}  J. Preskill, Proc. R. Soc. London A 454, 385 (1998).

\bibitem{9}  J. I. Cirac and P. Zoller, Phys. Rev. Lett. 74, 4091 (1995).

\bibitem{10}  C. Monroe et al., ibid 75, 4714 (1995).

\bibitem{11}  N. A. Gershenfeld and I. L. Chuang, Science 275, 350 (1997).

\bibitem{12}  I. L. Chuang et al., Nature 393, 143 (1998).

\bibitem{13}  D. G. Cory et al., quant-ph/9802018.

\bibitem{14}  B. E. Kane, Nature 393, 133 (1998); D. P. DiVincenzo, ibid
393, 113 (1998).

\bibitem{15}  C. P. Slichter, {\it Principles of Magnetic Resonance}, 3rd.
ed. (Springer-Verlag, New York, 1990).

\bibitem{16}  fL. Viola and S. Lloyd, quant-ph/9803057.

\bibitem{17}  D. P. DiVincenzo, Science 270, 255 (1995).

\bibitem{18}  B. Schumacher, Phys. Rev. A 54, 2614 (1996).

\bibitem{19}  L. M. Duan and G. C. Guo Phys. Rev. A 56, 4466 (1997).

\bibitem{20}  L. M. Duan and G. C. Guo Phys. Rev. Lett. 79, 1953 (1997).

\bibitem{21}  S. Lloyd, ibid 75, 346 (1995).

\bibitem{22}  W. S. Warren, N. A. Gershenfeld, and I. L. Chuang, Science
277, 1688 (1997).
\end{thebibliography}
\end{document}